\documentclass[10pt, journal,comsoc]{IEEEtran}
\usepackage[utf8]{inputenc}
\usepackage[T1]{fontenc}
\usepackage[english]{babel}
\usepackage{stackengine}

\ifCLASSOPTIONcompsoc
  \usepackage[nocompress]{cite}
\fi

\usepackage{fixltx2e}
\usepackage{graphicx}
\graphicspath{{./pics/}}
\usepackage{amsmath}
\usepackage{dsfont}				
\usepackage{url}
\usepackage{lipsum}
\usepackage{xcolor}
\usepackage{booktabs}
\usepackage{multirow}
\usepackage{makecell}
\usepackage{enumitem}

\usepackage[backend=bibtex,bibencoding=ascii,natbib,url=false, isbn=false,doi=true,maxbibnames=9,maxcitenames=3,style=numeric-comp,bibstyle=ieee]{biblatex}

\renewbibmacro{in:}{%
	\ifentrytype{inproceedings}{}
	{\printtext{\bibstring{in}\intitlepunct}}
	}

\makeatletter
\def\blx@maxline{77}
\makeatother

\definecolor{citblue}{rgb}{0.4,0.4,0.9999} 		
\makeatletter
\patchcmd{\@maketitle}
{\addvspace{0.5\baselineskip}\egroup}
{\addvspace{-1\baselineskip}\egroup}
{} {} \makeatother

\begin{document}
\title{A Multistage Method for SCMA Codebook Design Based on MDS Codes}

\author{Bruno Fontana da Silva,~\IEEEmembership{Student Member,~IEEE,}
	Danilo Silva,~\IEEEmembership{Member,~IEEE,} \\
	Bartolomeu F. Uchôa-Filho,~\IEEEmembership{Senior Member,~IEEE,}
	Didier Le Ruyet,~\IEEEmembership{Senior Member,~IEEE}
	\IEEEcompsocitemizethanks{\IEEEcompsocthanksitem 
		 This work was partially supported by {CNPq, Brazil, through PVE Project 400703/2014-9, SWE Program 204676/2018-5, and Ph.D. scholarship 141161/2016-7}. B. F. da Silva is with Instituto Federal Sul-rio-grandense (IFSul), Sapiranga, Brazil (e-mail: brunosilva@ifsul.edu.br). D. Silva and B. F. Uchôa-Filho are with the GPqCom/LCS/EEL,
		Federal University of Santa Catarina, Florianópolis 88040-900, Brazil (e-mails:
		$\lbrace $danilo, uchoa$ \rbrace$ @eel.ufsc.br). D. Le Ruyet is with CEDRIC, Conservatoire National des Arts et Metiers, 75003 Paris, France (e-mail: didier.le\_ruyet@cnam.fr).					
		}
}
%
%
\IEEEtitleabstractindextext{%
\begin{abstract}

Sparse Code Multiple Access (SCMA)  has been recently proposed for the future generation of wireless communication standards. SCMA system design involves specifying several parameters. In order to simplify the procedure, most works consider a multistage design approach. Two main stages are usually emphasized in these methods: sparse signatures design (equivalently, resource allocation) and codebook design. In this paper, we present a novel SCMA codebook design method. 
The proposed method 
considers SCMA codebooks structured
with an underlying vector space obtained from classical block codes. In particular, when using maximum distance separable (MDS) codes, our proposed design provides maximum signal-space diversity  with a relatively small alphabet. The use of small alphabets also helps to maintain desired properties in the codebooks, such as low peak-to-average power ratio and low-complexity detection.
\end{abstract}
\begin{IEEEkeywords}
	SCMA, NOMA, MDS codes, multidimensional modulations, signal-space diversity.
\end{IEEEkeywords}}
\maketitle
\IEEEdisplaynontitleabstractindextext
\section{Introduction}

{Sparse Code Multiple Access (SCMA) is a Code-Domain Non-Orthogonal Multiple Access (CD-NOMA) first proposed in \cite{nikopour_sparse_2013}. SCMA has been recently considered as a potential candidate for multiple access techniques of the fifth generation of wireless communication standards (5G). Designing a SCMA system involves several parameters (e.g. sparse signatures, codebook design and the multiuser detector). Hence, several works \cite{nikopour_sparse_2013,taherzadeh_scma_2014,bao_bit-interleaved_2018} consider suboptimal approaches using multistage design methods in order to simplify the design steps. }

{In multistage design methods for SCMA, the first step usually implies resource allocation, i.e. designing sparse spreading signatures for the users. The sparsity causes the number of collisions in each resource to be small compared to the total number of users in the system,  
enabling joint multiuser detection (MUD) with relatively low complexity using the message-passing algorithm (MPA) \cite{nikopour_sparse_2013, bayesteh_low_2015}.}

{Although many papers \cite{nikopour_sparse_2013,xiao_capacity-based_2018,bao_bit-interleaved_2018} present benchmarks with very small and dense resource allocation matrices, 
the sparse structure initially proposed for SCMA, 
similar to Low-Density Parity Check (LDPC) codes, 
suggests that the MPA receiver works better in larger dimensions. 
Recent works such as \cite{he_nonbinary_2017,li_high-dimensional_2018} proposed designing the SCMA resource allocation matrices using design methods of LDPC codes, achieving better performance with larger dimension (less dense matrices) systems.}

{The second design step involves codebook design on top of the spreading signatures. The work in \cite{nikopour_sparse_2013} established SCMA transmitters as multidimensional modulation (MDM) schemes where each user maps a sequence of bits directly to a codeword in an $N$-dimensional finite set of $M$ complex codewords, which is denoted as a SCMA codebook. These codewords are further spread over $K$ shared resources using the 
corresponding
sparse signature vector designed for each user.}

{Authors in \cite{taherzadeh_scma_2014} proposed guidelines to design 
SCMA codebooks based on 
a mother-constellation as the baseline to all users' codebooks. The work \cite{bayesteh_low_2015} proposed codebooks with a low number of complex projections (LNCP) per complex dimension, which are able to simultaneously reduce the peak-to-average power ratio (PAPR) of the modulation and the MPA receiver complexity. In \cite{bao_bit-interleaved_2018}, LNCP codebooks are designed based on Amplitude and Phase Shift Keying (APSK) one-dimensional modulations and then extended to $N$-dimensional MDMs by searching for permutation patterns. In \cite{xiao_capacity-based_2018}, the codebooks are designed solving optimization problems derived from approximations of the constrained-constellation capacity of one dimensional multiple-access channels, also extending it to $N$-dimensional MDMs with permutation searches.}

{In this paper, we consider a multistage design for an uplink SCMA system and we design $M$-ary $N$-dimensional codebooks based on block codes over $q$-ary alphabets. The 
benefits 
of our proposed method are the following:
\begin{enumerate}[label=(\alph*)]
	\item  in the codebook design procedure, we can use off-the-shelf block codes and avoid exhaustive permutation searches found in other design methods;
	\item by properly choosing $q$, we can maintain in our codebooks the low PAPR and  low-complexity MUD properties described in the methods given in \cite{bayesteh_low_2015,bao_bit-interleaved_2018};
	\item when using maximum distance separable (MDS) codes, our method has an additional benefit of guaranteeing arbitrarily large minimum signal-space diversity (MSSD).
\end{enumerate}
In addition, 
we exploit large size SCMA systems in this paper, a scenario that has been given scant attention in the literature. We show design examples of SCMA codebooks and numerical results in which they achieve better performance when comparing with codebooks based on other known approaches. }

\section{System Model} \label{sec_SysModel}

{We consider the uplink of $J$ users over $K$ shared orthogonal resources. The load factor is defined as $J/K$ and the system is said to be overloaded if $J>K$, which is the scope of SCMA. The uplink channel $\mathbf{h}_j \in \mathds{C}^K$ of user $j$ is modeled with independent Rayleigh fading coefficients, i.e. $h_j(k) \sim \mathcal{CN}\left(0,1\right)$. Over the uplink Rayleigh fading channel, we can assign the same SCMA codebook to each user without the need of individual operators such as phase-offset between users \cite{vameghestahbanati_multidimensional_2019}. Hence, each user maps its $b$ information bits to an $N$-dimensional  codeword $\mathbf{x}$ of an $M$-ary codebook $\mathcal{X} \subset \mathds{C}^N$. 
Finally, 
individual $K \times N$ binary mapping matrices $\mathbf{V}_j$ spread each user's $N$-dimensional symbol over the $K$ resources. The system model is written as
\begin{equation} \label{eq_sysmodel}
	\mathbf{y} = \sum_{j=1}^{J} \mathbf{H}_j \mathbf{x}_j  + \mathbf{z},
\end{equation}
where  $\mathbf{H}_j=\text{diag}\left( \mathbf{h}_j\right) \mathbf{V}_j$ is the equivalent channel of user $j$ and $\mathbf{z} \in \mathds{C}^K$ is the noise vector whose entries are modeled as additive white Gaussian noise (AWGN), i.e. $z(k)\sim \mathcal{CN} \left(0, \sigma_z^2 \right)$.}

\begin{figure}[t]
	\centering
	\includegraphics[width=.60\linewidth]{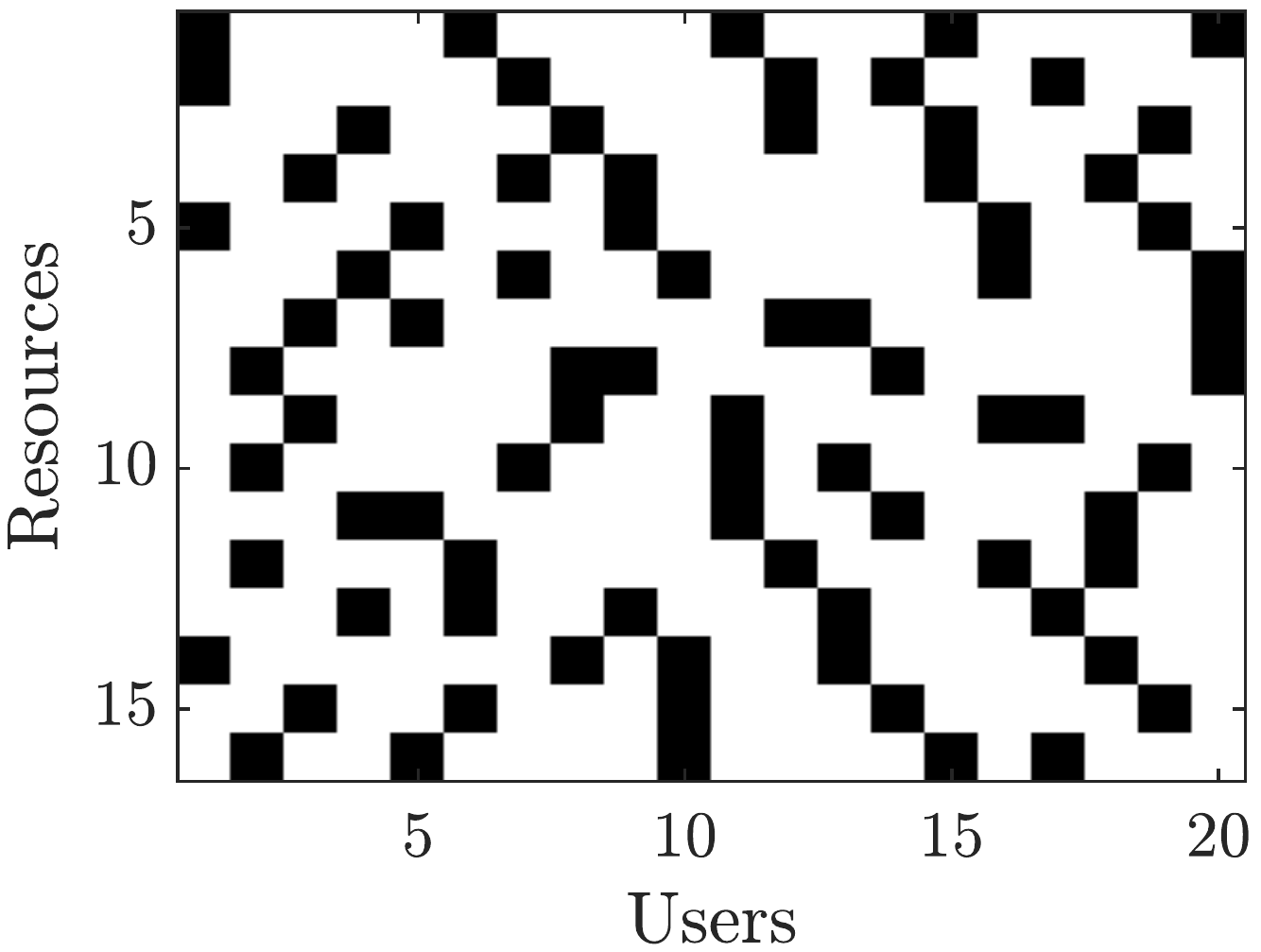}
	\caption{{Design example of the resource allocation matrix $\mathbf{F}$ for $J=20$ user and $K=16$ resources. Black squares represent entries $\mathbf{F}[j,k]=1$, corresponding to resource $k$ allocated to user $j$. White spaces represent null entries, meaning that the user does not spread over this resource.} 
	}
	\label{fig:bipartitegraphoff}
\end{figure}

{The mapping matrices $\mathbf{V}_j$ can be easily represented in a  $K \times J$ binary resource allocation matrix $\mathbf{F}$, where the columns correspond to the spreading signature of each user \cite{taherzadeh_scma_2014}. Without loss of generality, in this paper we assume that $\mathbf{F}$ is regular, i.e. it has constant column degrees $N$ and row degrees $d_f$.
Hence, the density of $\mathbf{F}$ is given as 
${N}/{K}$.
In Figure \ref{fig:bipartitegraphoff} we see a design illustration of a regular $\mathbf{F}$ for a SCMA system with $125\%$ overload, $N=4$ dimensional codebooks  and $d_f = 5$. }

\section{Multistage Design of the Codebook} \label{sec_MultiStageMethod}

{In this section we follow a multistage approach to design the uplink SCMA system modeled in Section \ref{sec_SysModel}. Given the desired parameters  $(J,K,N,M)$, the design framework consists 
of 
four sequential stages: (1) resource allocation design (matrix $\mathbf{F}$); (2) design of the block code codebook  structure; (3) codebook complex projections design and (4) binary labeling $\mu: \mathds{F}_2^b\rightarrow \mathcal{X}$ of the designed codebook. At the receiver side we assume the use of MPA for LNCP codebooks 
as described in \cite{bayesteh_low_2015}.}

{Our main contribution is related to the block code codebook structure in Stage (2). Using the block code structure we are able to design codebooks with 
large MSSD and maintain relatively low receiver complexity due to the use of small alphabets. We detail the procedure in the following sections.
}

\subsection{Stage (1): Resource Allocation Design} \label{stage1}
 {Most papers on SCMA \cite{nikopour_sparse_2013,bao_bit-interleaved_2018,xiao_capacity-based_2018} are restricted to the analysis of a low dimension $4 \times 6$ matrix and codebooks with $N=2$ dimensions. However, as pointed out in \cite{he_nonbinary_2017,li_high-dimensional_2018}, the performance of the MPA receiver can be improved when using matrices with larger girth, which is not possible in very dense (low dimensional) matrices. Moreover, to the best of our knowledge, no other work has considered feasible codebooks with $N>2$, limiting the MSSD $L$ of the MDM to be $1 \leq L\leq 2$. Since 
  we want to improve the MPA performance, with
  large dimensional codebooks, we consider matrices with $N>2$ and we design $\mathbf{F}$ with larger dimensions 
  in order to find larger girth and lower density matrices. 
  We 
  choose to 
  use 
  Progressive Edge-Growth (PEG) \cite{hu_regular_2005} algorithm in order to find a $K \times J$ binary matrix $\mathbf{F}$ with regular column-degree $N$.
  The impact of these design guidelines will be evaluated in Section \ref{sec_Results}. A design example can be seen in Figure \ref{fig:bipartitegraphoff} for input parameters $(J,K,N)=(20,16,4)$.}

\subsection{Stage (2): Proposed Block Code Structure for the MDM}  \label{stage2}
{
Most of the noteworthy SCMA codebooks rely on
a MDM with good properties for single-user scenarios \cite{taherzadeh_scma_2014,bao_bit-interleaved_2018}. One important figure of merit in MDMs is the MSSD $L$, which 
herein is associated with 
the minimum Hamming distance of a block code. Motivated by the MSSD figure of merit, we propose to design the SCMA codebook using the underlying structure of  well-known error-correcting block codes,
in particular MDS codes. To the best of our knowledge, no other work has considered such an approach for SCMA codebooks, which completely eliminates 
exhaustive search of permutations in \cite{bao_bit-interleaved_2018}.}

Consider a $q$-ary finite field $\mathds{F}_q=\left\lbrace \alpha_1,\alpha_2,\ldots,\alpha_{q}  \right\rbrace$, 
{where $q$ is a power of prime}.
A linear block code is the result of a one-to-one mapping between each vector $\mathbf{u}$ of a message-space $\mathds{F}_q^{k}$ and the elements $\mathbf{v} $ of a vector-subspace $\mathcal{V} \subset \mathds{F}_q^{N}$ generated by the generator matrix $\mathbf{G} \in \mathds{F}_q^{k \times N}$.  Both the message-space and the generated vector-space have size $q^{k}$. 

We propose that an $N$-dimensional $M$-ary SCMA codebook $\mathcal{X}$ {be} structured using the vector-subspace generated by $\mathbf{G}$ by associating the field elements 
$ \alpha_1,\ldots,\alpha_q  $ 
with the set 
$\mathcal{A}_q=\left\lbrace x_1, \ldots, x_q \right\rbrace$, where ${x}_i \in \mathds{C}$ for $i=1,\ldots,q$.

If the generator matrix of a MDS linear code is chosen, the resulting minimum Hamming distance of the linear code will be $d_{\rm mds} = N-k +1 $. This result translates to a {MSSD} $L=d_{\rm mds}$ in the SCMA codebook $\mathcal{X}$.  {Since $k=1$ usually corresponds to repetition codes}, we consider SCMA codebooks based on MDS block codes {with} $k \geq 2$.

In summary, (a) {choose} $k$ and $q$, considering that $|\mathcal{X}|=M\leq q^{k}$; (b) {choose} a generator matrix $\mathbf{G}_{k \times N}$ of an MDS code, (c) use $\mathbf{G}$ to generate 
a $(N,k,d_{\rm mds})_q$ MDS code,
(d) in each of the 
the codewords, replace the field elements 
$ \alpha_1,\ldots,\alpha_q $ 
by the elements $x_1,\ldots,x_q$, respectively. 

\textbf{Design Example 1}: in this example, we use a Generalized Reed-Solomon (GRS) construction 
in order to achieve length $N=q$. Consider a $\left(N, k \right)=\left(4, 2 \right)$ code {so} that $d_{\rm mds}=N-1=3$ and we can use alphabet size $q=4$. This defines the maximum size of the codebook, $M \leq 16$, and the {MSSD} $L=3$. We fix $M=16$ and use all the codewords in order to have a $16$-ary MDM.  The $k \times N$ generator matrix of this GRS,
in 
systematic form, is given by 
$\mathbf{G}_{2 \times 4} = \begin{bmatrix}
\mathbf{I}_2 & \begin{bmatrix}
\alpha & \alpha^2
\end{bmatrix}^T & \begin{bmatrix}
\alpha^2 & \alpha 
\end{bmatrix}^T
\end{bmatrix}$,
which spans $16$ vectors {in $\mathds{F}_4^4$}. We replace the field elements 
$  0,1,\alpha,\alpha^2 $
 by the respective {complex} elements of $\mathcal{A}_q=\left\lbrace x_1, x_2, x_3,x_4 \right\rbrace$ in each dimension. At this point, 
$\mathcal{X}$ is structured as
\begin{equation} \label{block_code_natural}
\resizebox{.91\linewidth}{!}{
	$
	\begin{bmatrix}
	x_1&x_3&x_4&x_2&x_1&x_3&x_4&x_2&x_1&x_3&x_4&x_2&x_1&x_3&x_4&x_2 \\
	x_1&x_1&x_1&x_1&x_3&x_3&x_3&x_3&x_4&x_4&x_4&x_4&x_2&x_2&x_2&x_2 \\
	x_1&x_4&x_2&x_3&x_2&x_3&x_1&x_4&x_3&x_2&x_4&x_1&x_4&x_1&x_3&x_2 \\
	x_1&x_2&x_3&x_4&x_4&x_3&x_2&x_1&x_2&x_1&x_4&x_3&x_3&x_4&x_1&x_2
	\end{bmatrix}.
	$
}
\end{equation}

\textbf{Design Example 2}: consider the codebook design for $M=8$ codewords and spreading degree $N=4$. We set $q=3$ and $k=2$. Using a Hamming code construction for $\left(N,k,d\right)_q = \left(4,2,3 \right)_3$, we obtain an MDS code with $d_{\rm mds}=3$. The $k \times N$ generator matrix
in systematic form, given by $\mathbf{G}_{2 \times 4} = \begin{bmatrix}
\mathbf{I}_2 & \begin{bmatrix}
1 & 1
\end{bmatrix}^T & \begin{bmatrix}
1 & 2
\end{bmatrix}^T
\end{bmatrix}$, is able to span a $9$-ary 
vector-{subspace} of $\mathds{F}_3$ which is actually an MDS code.  We denote this design as the Hamming Code (HC) construction. In order to obtain a $8$-ary codebook we must expurgate one codeword. This procedure will be done using figures of merit detailed in the following Stage (3). 


\subsection{Stage (3): Designing the Complex Projections} \label{stage3}

Choosing the $q$ projections is still necessary in Stage (3). In \cite{bao_bit-interleaved_2018}, $\mathcal{A}_q$ is arbitrarily chosen as an APSK constellation as the first step of codebook design. We relax the problem for the choice {of} any other set using any desired optimization method over the proposed structure.  Since the MPA receiver calculates the messages independently in each resource, the codebooks could also have a different set $\mathcal{A}_q$ in each of the $N$ dimensions, i.e., $\mathcal{A}_q^{(n)}$ for $n=1,\ldots,N$. The choice of the $N$ sets $\mathcal{A}_{q}^{(n)}$ can optimize a codebook figure of merit of interest. For example, the mean cutoff rate approximation $\Psi \left( \mathcal{X} \right)$ suggested in \cite{bao_design_2016} can be used in the search {for the $q N$ projections}. 
Other figures of merit to be considered are the minimum Euclidean distance and the minimum product distance. The previously designed {MSSD} $L$ is not affected as long as the $q$ projections have no multiplicity, i.e. $x_i\neq x_j$ for all $i\neq j$. 

During simulation analysis, we observed that optimizations over $\Psi \left( \mathcal{X}\right)$ do not improve significantly the performance of the system when comparing with arbitrary choices for $\mathcal{A}_q$ such as APSK projections, especially for low $q$ such as $3$ or $4$. Thus, in this letter, we restrict our analysis for {LNCP codebooks} with small values of $q$ and we simplify $\mathcal{A}_{q}^{(n)}=\mathcal{A}_{q}$ $\forall n$ by choosing arbitrary $q$ complex projections among known constellations such as QAM and APSK.

On the other hand, for cases {where} $q^{k}$ is greater than the desired $M$ and once $\mathcal{A}_q$ is fixed, it is still possible to expurgate $q^{k} - M$ vectors in order to optimize some figure of merit dependent on $\mathcal{X}$, e.g. maximizing $\Psi\left( \mathcal{X} \right)$. {For the HC in the second design example of Section \ref{stage2}, we optimized the MDM over $\Psi\left(\mathcal{X} \right)$ with APSK projections by removing the HC codeword {produced by the message vector} $\mathbf{u}=\left(2,2\right)$.}

\subsection{Stage (4):  Designing the Binary Labeling} \label{stage4}

Once the codebook $\mathcal{X}$ is fully designed from Stages (1) to (3), the symbol-error rate (SER) of the system is defined for a given MUD. Yet one can perform optimization by choosing a binary labeling function $\mu$ among the $M!$ possible solutions in order to improve the resulting bit error rate (BER). 
For classical one-dimensional (complex) rectangular QAM constellations, Gray-code labeling provides a reduction factor of $1/k$ from the SER to the BER curve. For {general MDMs}, the binary-switching algorithm (BSA) has been proposed in \cite{schreckenbach_optimization_2003} in order to find a good labeling by optimizing a cost function based on the Euclidean distance between codeword pairs. 
In this paper we also consider the BSA approach.

\subsection{Complexity reduction in the codebook design stage}

{
To extend the design from $1$ to $N$ dimensions,  Step 2 of the codebook design method in \cite{bao_bit-interleaved_2018} proposes exhaustive search of integer permutations, which has complexity $\mathcal{O} \left((M!)^{N-1}\right)$, intractable even for small values of $M$ and $N$ such as $M=8$, $N=3$. Even the suboptimal approach in \cite{bao_bit-interleaved_2018} has complexity $\mathcal{O} \left( \left(N-1\right)\times M! \right) $ which is also intractable for high order constellations. In \cite{xiao_capacity-based_2018}, the codebook design involves evaluating a cost function with complexity $M^{d_f}$ (same complexity as the original MPA receiver), which clearly  is  not a practical design as  $M$ and $d_f$ increase; additionally, in a second step of their method, they also face the permutation search problem of \cite{bao_bit-interleaved_2018}.}

In our Stage (2) we obtained a set of integer permutations in a straightforward manner using $\mathbf{G}_{k \times N}^T$ and linear operations to span a vector space. The complexity order of this procedure is $\mathcal{O}\left( M \right)$. Hence, our proposed method avoids large complexity during the codebook design stage and enables a straightforward design of high-order constellations with the extra benefit of ensuring arbitrary MSSD up to the MDS bound. If desired, one may also use non-MDS codes, which limits the MSSD to $L<N-k+1$ but may relax some design constraints. 

\subsection{{On the MPA receiver complexity}}

{Regardless of the codebook design method, when using the MPA as MUD, the receiver complexity is still exponential. However, using $q < M$ complex projections per dimension in the codebook can significantly reduce the complexity of the MPA receiver \cite{bayesteh_low_2015}. The receiver complexity of the original MPA for SCMA {is $\mathcal{O} \left(K d_f M^{d_f} \right)$~\cite{bayesteh_low_2015}. With LNCP codebooks, this complexity order becomes   $\mathcal{O}\left(K d_f q^{d_f}\right)$ ($q\leq M$).} In this sense, by using small alphabet size $q<M$ in our proposed Stage (2), we maintain the benefits of LNCP codebooks proposed in \cite{bayesteh_low_2015}. 
}
\section{Simulation Results} \label{sec_Results}

In this section, we present Monte Carlo simulations of the uncoded BER 
for the proposed method over {the independent,  Rayleigh fading channel model given in~\eqref{eq_sysmodel}.} 
The low-complexity $\log$-MPA receiver based on \cite{bayesteh_low_2015} is used with $5$ iterations. BSA was applied with the same number of iterations as a last step optimization for all the designed constellations.  For clarity, we define $\frac{E_b}{N_0} = \frac{\rm SNR }{\log_2{M}}$. 

{To the best of our knowledge, there are no published SCMA schemes with $N=4$ and tractable decoding complexity}. {Since the codebooks in \cite{bao_bit-interleaved_2018} have outperformed the ones in \cite{taherzadeh_scma_2014}, and since we consider that the design complexity of \cite{xiao_capacity-based_2018} is not practical, even for moderate $M$ and $d_f$,} we compare our proposed MDMs with 
$2$
different codebooks that we have designed following the method from \cite{bao_bit-interleaved_2018}.  We have considered APSK based $A_{M,q}$ for $\left(M,q \right)=\left\lbrace \left( 8, 3\right), \left(16, 4\right) \right\rbrace $. Since exhaustive search of the permutations is not tractable, we considered the suboptimal approach proposed for the step 2 in \cite{bao_bit-interleaved_2018} and we searched the permutations randomly over $T=10^5$ iterations per dimension (complexity {$\mathcal{O}\left(\left(N-1\right)T\right)$}). 
Both the proposed and 
benchmark 
codebooks consider APSK projections which are distributed over the rings according to vectors $\mathbf{m}=[3]$ and $\mathbf{m}=[4]$ for $q=3$ and $q=4$, respectively  
(see more details in \cite{bao_bit-interleaved_2018}). For the proposed codebooks, we reuse our design examples from \ref{stage2} based on {the} HC and GRS constructions. In Table \ref{codebook_params}, we provide values of some figures of merit of 
these
codebooks.

\begin{figure}[]
	\centering
	\hspace*{-3mm}
%
	\includegraphics[width=3.00in]{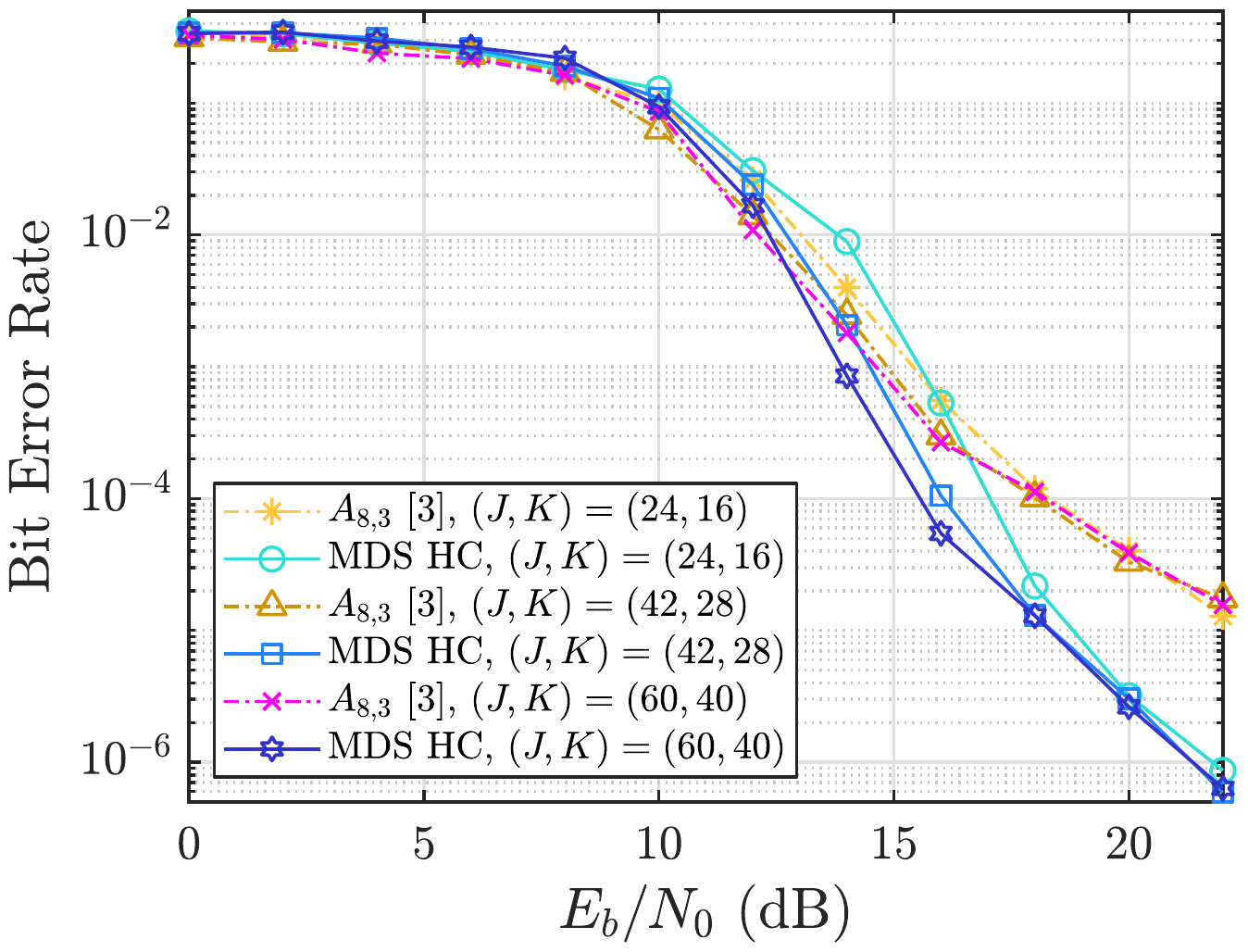}
	\caption{{MPA performance for different density levels of the resource allocation matrices $\mathbf{F}$. The  load is fixed to {$J/K=150 \%$ ($d_f=6$)}, and the codebook structure is the HC-based one, with $\left( N,k,M,q \right)= \left( 4, 2, 8, 3 \right)$.}}
	\label{fig:letter2019ebn0dbxerrorratejkcomparisonov120q3}
	\vspace{-4mm}
\end{figure}

\begin{figure}[]
	\centering
	\hspace*{-3mm}
	\includegraphics[width=3.00in]{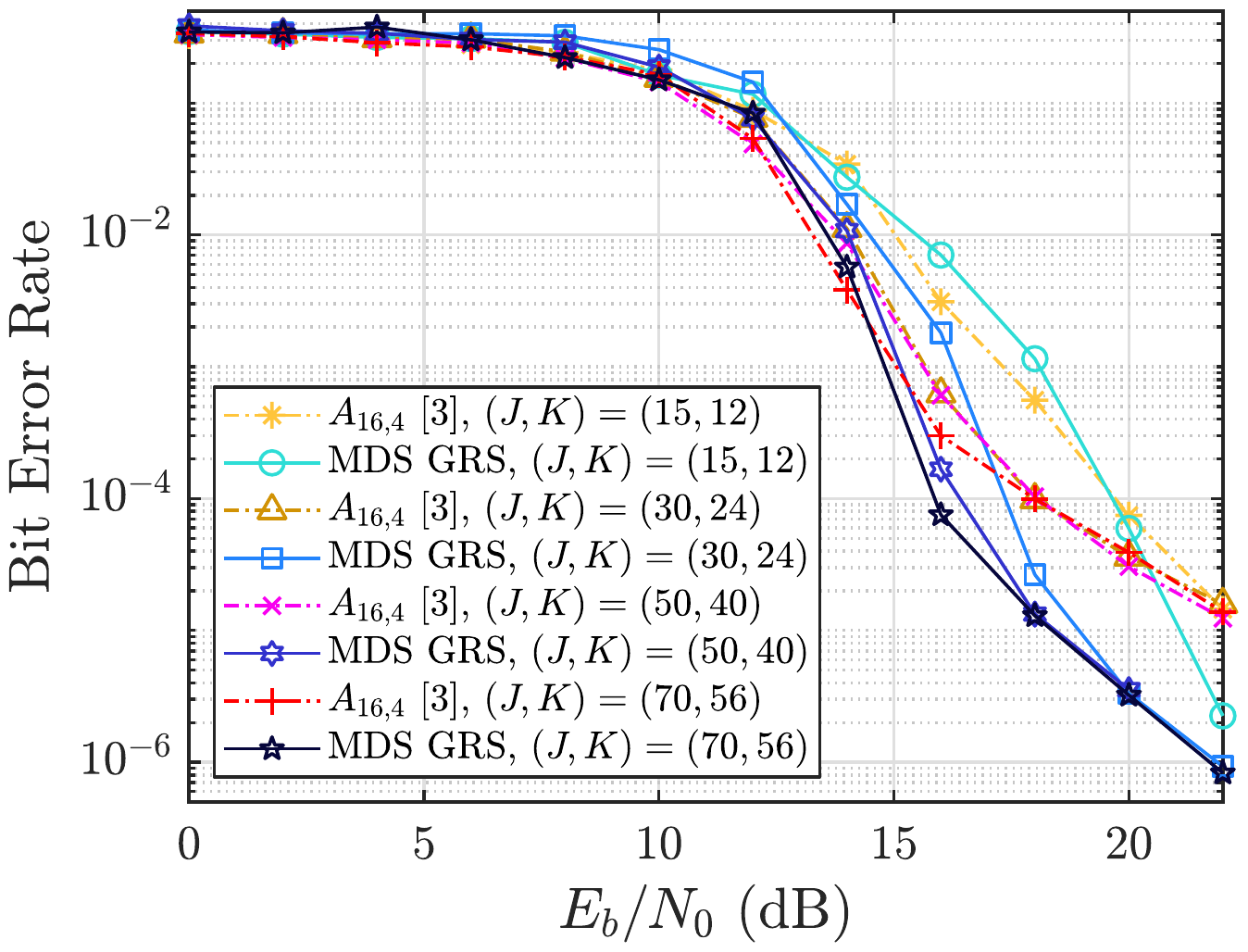}
	\caption{{MPA performance for different density levels of the resource allocation matrices $\mathbf{F}$. The load is fixed to {$J/K=125 \% $ ($d_f=5$)}, and the codebook structure is the GRS-based one, with $\left( N,k,M,q \right)= \left( 4, 2, 16, 4 \right)$.}}
	\label{fig:letter2019ebn0dbxerrorratejkcomparisonov120q4}
	\vspace{-4mm}
\end{figure}

{Figures }\ref{fig:letter2019ebn0dbxerrorratejkcomparisonov120q3} and \ref{fig:letter2019ebn0dbxerrorratejkcomparisonov120q4} show the MPA performance according to the output matrices $\mathbf{F}$ from the procedure {in} Stage (1) for several combinations of $(J,K)$, which are indicated in the legends. The matrices were designed with fixed overload: 
{$J/K=150\%$ in Figure \ref{fig:letter2019ebn0dbxerrorratejkcomparisonov120q3}  
and 
$J/K=125\%$ in Figure \ref{fig:letter2019ebn0dbxerrorratejkcomparisonov120q4}}. 
%
%
The {smallest matrix has a density of {$33\%$}}, while the largest {one} has a density of {$7.1\%$}. 
The simulation results show that reducing the density of $\mathbf{F}$ by scaling $J$ and $K$ up to a certain size provides {significant performance improvement},  {especially for larger $M$}. 

Overall, simulation results show that {the MDS-based codebooks, with larger MSSD, yield better performance.} Since the receiver complexity is proportional to $q^{d_f}$, there is a trade-off when choosing $q=4$ for larger diversity, although the data rate can increase proportionally to $q$. 
In the web page \cite{noauthor_supplementary_2018} we provide supplementary material with 
files for the simulated codebooks and the designed 
resource allocation matrices. 
\begin{table}[]
	\centering
	\caption{Figures of merit comparison for $16$-ary $4$-dim. codebooks.}
	\label{codebook_params}
		\begin{tabular}{|c|c|l|l|l|l|}
			\hline
			\makecell {Design \\ Methods  } & $\left(M, q, N \right)$ & $\min d_E^2$ & $\min d_{p,L}$  & $L$ & \makecell{$\Psi\left(\mathcal{X} \right)$ \\ @ $8$ dB} \\ \hline
			\multirow{2}{*}{\makecell{ Proposed \\ Codebooks}} & $\left( 8, 3, 4 \right)$ & $2{.}25$ & $0{.}6495$  & $3$  & $2{.}2864$ \\ \cline{2-6} 
			&  $\left( 16, 4, 4 \right)$  & $2$  & $0{.}5$ & $3$ & $3{.}0167 $ \\ \cline{2-6} \hline
			\multirow{2}{*}{\makecell{ Method \\ from  \cite{bao_bit-interleaved_2018}}} & $\left( 8, 3, 4 \right)$ & $1{.}5$  & $0{.}75$ & $2$ & $2{.}2337$ \\ \cline{2-6} 
			& $\left( 16, 4, 4 \right)$ & $1$ & $0{.}5$  & $2$  & $2{.}9548$  \\ \cline{2-6} \hline
		\end{tabular}
	\vspace{-3mm}
\end{table} 	

\section{Conclusion}

In this paper, we proposed a novel SCMA codebook design method. {We revisited the multistage approaches from the current literature,} incorporating trade-offs between performance and complexity and allowing designs for a wide and scalable range of system parameters. {Our codebook design method} has linear complexity and algebraic properties, when compared to the exponential complexity of permutation searches {of} other methods. The MDS block codes structure provides a new direction for a new family of SCMA codebooks based on classical code designs.

\AtNextBibliography{\scriptsize}
\printbibliography[heading=bibnumbered]

\end{document}